\documentclass{ckm}                 % twocolumn proceedings style
\usepackage{txfonts}            % if you have it, to get Times family
\def\DAF{DA\char8NE}  
\newcommand{\kl}{\mbox{$K_L$}}
\newcommand{\ks}{\mbox{$K_S$}}
\newcommand{\ksl}{\mbox{$K_{S,L}$}}
\newcommand{\kp}{\mbox{$K^+$}}
\newcommand{\km}{\mbox{$K^-$}}
\newcommand{\kpm}{\mbox{$K^\pm$}}

\newcommand{\DKSpippim}{\mbox{$K_S\rightarrow\pi^+\pi^-$}}
\newcommand{\DKSeIII}{\mbox{$K_S\rightarrow\pi e \nu$}}
\newcommand{\kppo}{\mbox{$K^{\pm}\rightarrow\pi^{\pm}\pi^{0}$}}
\newcommand{\kmunu}{\mbox{$K^{\pm}\rightarrow\mu^{\pm}\nu$}}
\newcommand{\kppoo}{\mbox{$K^{\pm}\rightarrow\pi^{\pm}\pi^{0}\pi^{0}$}}
\newcommand{\eV}{{e\kern-.07em V}}
\newcommand{\MeV}{{\rm \,M\eV}}
\newcommand{\GeV}{{\rm G\eV}}
\newcommand{\ps}{{\rm \,ps}}
\newcommand{\ns}{{\rm \,ns}}
\newcommand{\mm}{{\rm \,mm}}
\newcommand{\cm}{{\rm \,cm}}
\newcommand{\m}{{\rm \,m}}
\newcommand{\um}{\ensuremath{\mathrm{\mu m}}}
\newcommand{\T}{{\rm \,T}}
\newcommand{\Lpb}{\ensuremath{\rm pb^{-1}}}
\newcommand{\tzero}{\ensuremath{T_{0}}}

\newcommand{\Vus}{\ensuremath{\mathrm{V_{us}}}}
\newcommand{\Vud}{\ensuremath{\mathrm{V_{ud}}}}
\newcommand{\Obs}{\ensuremath{\mathrm{|f^{K \pi}_{+}(0)\cdot V_{us}|}}}
\newcommand{\Keltre}{\ensuremath{\mathrm{K_{\ell 3}}}}
\newcommand{\Kpmetre}{\ensuremath{\mathrm{K^{\pm}_{e3}}}}
\newcommand{\Kpmmutre}{\ensuremath{\mathrm{K^{\pm}_{\mu3}}}}

\newcommand{\KOetre}{\ensuremath{\mathrm{K^{0}_{e3}}}}
\newcommand{\KOmutre}{\ensuremath{\mathrm{K^{0}_{\mu3}}}}
\newcommand{\fpiuo}{\ensuremath{f_{+}(0)}}
\newcommand{\ff}{\ensuremath{f_{\pm,0}(t)}}
\newcommand{\taul}{\ensuremath{\tau_{K_L}}}
\newcommand{\taupm}{\ensuremath{\tau_{K^\pm}}}
\newcommand{\lp}{\ensuremath{\lambda_{+}}}
\newcommand{\lz}{\ensuremath{\lambda_{0}}}

\confname{Workshop on the CKM Unitarity Triangle, IPPP Durham, April
  2003}

\title{KLOE prospects and preliminary results for \Keltre\  decay measurements}

\author{\centerline{The KLOE Collaboration\thanks{
A.~Aloisio, F.~Ambrosino, A.~Antonelli, M.~Antonelli, C.~Bacci, G.~Bencivenni,
S.~Bertolucci, C.~Bini, C.~Bloise, V.~Bocci, F.~Bossi, P.~Branchini, S.~A.~Bulychjov,
R.~Caloi, P.~Campana, G.~Capon, T.~Capussela, G.~Carboni, G.~Cataldi, F.~Ceradini,
F.~Cervelli, F.~Cevenini, G.~Chiefari, P.~Ciambrone, S.~Conetti, E.~De~Lucia,
P.~De~Simone, G.~De~Zorzi, S.~Dell'Agnello, A.~Denig, A.~Di~Domenico, C.~Di~Donato,
S.~Di~Falco, B.~Di~Micco, A.~Doria, M.~Dreucci, O.~Erriquez, A.~Farilla, G.~Felici,
A.~Ferrari, M.~L.~Ferrer, G.~Finocchiaro, C.~Forti, A.~Franceschi, P.~Franzini, C.~Gatti,
P.~Gauzzi, S.~Giovannella, E.~Gorini, E.~Graziani, M.~Incagli, W.~Kluge, V.~Kulikov,
F.~Lacava, G.~Lanfranchi, J.~Lee-Franzini, D.~Leone, F.~Lu, M.~Martemianov, M.~Matsyuk,
W.~Mei, L.~Merola, R.~Messi, S.~Miscetti, M.~Moulson, S.~M\"uller, F.~Murtas, M.~Napolitano,
A.~Nedosekin, F.~Nguyen, M.~Palutan, E.~Pasqualucci, L.~Passalacqua, A.~Passeri, V.~Patera,
F.~Perfetto, E.~Petrolo, L.~Pontecorvo, M.~Primavera, F.~Ruggieri, P.~Santangelo,
E.~Santovetti, G.~Saracino, R.~D.~Schamberger,
{\bf B.~Sciascia  (corresponding author: Laboratori Nazionali di Frascati dell'INFN, 
via E.Fermi, 44 00044 Frascati, Italy)},
A.~Sciubba, F.~Scuri, I.~Sfiligoi, A.~Sibidanov, T.~Spadaro, E.~Spiriti, M.~Testa,
L.~Tortora, P.~Valente, B.~Valeriani, G.~Venanzoni, S.~Veneziano, A.~Ventura,
S.~Ventura, R.~Versaci, I.~Villella, Y.~Xu.
}
}}

\begin{document}

\begin{abstract}
  The KLOE experiment has been running since April 1999 at the \DAF\ 
  e$^{+}$-e$^{-}$ collider at a center of mass energy centered around 
  the $\phi$ meson mass.  The luminosity integrated up to September 2002
  is $\sim 500~$\Lpb . Here we present the perspectives on the  
  measurement of the $V_{us}$ CKM-matrix element with the KLOE detector, 
  using both charged and neutral kaon semileptonic decays.
\end{abstract}

%% \maketitle needs to be after the author and address info and the abstract... 
\maketitle
%% standard LaTeX from here on...

At present, determinations of \Vus\ and \Vud\ provide the most precise constraints
on the size of CKM matrix elements; in particular the most accurate determination of \Vus\ is
obtained from semileptonic decays of both neutral and charged kaons (\Keltre). 
Concerning the steps necessary to extract \Vus\ from 
the experimental determination of \Keltre\ decay rates,
specifically the theoretical evaluation of \fpiuo\  
and the theoretical treatment of photonic radiative corrections,
we refer the reader to the contribution to these proceedings 
by V.~Cirigliano~\cite{Ckm2003} and 
to the proceedings of the previous {\it CKM Workshop}~\cite{CKMyellow2002}.

Measurements of the branching ratios and the momentum dependence of \ff\ (\lp\ and \lz)
for the decays \Kpmetre , \Kpmmutre , \KOetre , and \KOmutre ,
together with the lifetimes \taul\ and \taupm , 
allows four independent 
determinations of the observable \Obs\ to be obtained.
At KLOE, we have the possibility of measuring the full set of kaon
semileptonic decays using the same detector; 
moreover, in all channels we can use the tagging technique.

We will start with a summary of the characteristics of the KLOE detector and of the data set.
We will then describe the status of the 
different contributions that the KLOE experiment 
can make in the field of \Vus\ determination.
\section*{The KLOE experiment at \DAF}
KLOE~\cite{kp,tp} operates at \DAF~\cite{dafne}, an
$e^+e^-$ collider also known as the Frascati $\phi$ factory; 
$\phi$ mesons are produced in small angle (25 mrad) collisions of
equal energy electrons and positrons, giving the $\phi$ a small transverse momentum component
in the horizontal plane, $p_\phi\!\sim\!13\MeV/c.$
The main advantage of studying kaons at a $\phi$ factory is that $\phi$ mesons decay
$\sim\!49\%$ of the time into charged kaons and
$\sim\!34\%$ of the time into neutral kaons. \kl's and \ks's (or \kp's and \km's) 
are produced almost back-to-back in the laboratory, 
with mean decay paths $\lambda_{\rm L}\!\sim\!340$\cm , 
$\lambda_{\rm S}\!\sim\!0.6$\cm, and $\lambda_{\rm \pm}\!\sim\!90$\cm, 
respectively. 
One of the features of a $\phi$ factory is the
the possibility to perform tagged measurements: the detection
of a  long-lived neutral kaon \kl\ guarantees the presence of a \ks\ of given
momentum and direction  and vice versa. The same holds for charged kaons.
\begin{figure}
  \hbox to\hsize{\hss
    \includegraphics[width=\hsize]{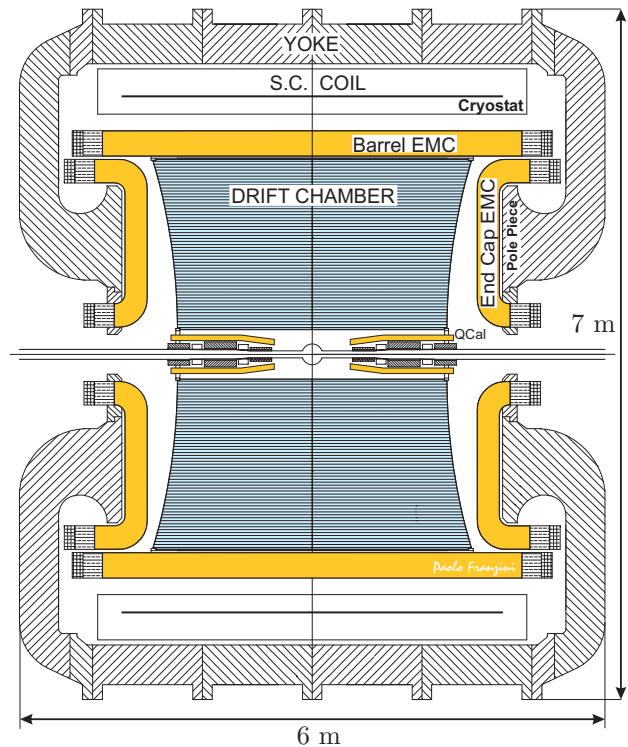}
    \hss}
  \caption{Vertical cross section of the KLOE detector.}
  \label{fig:kloesec}
\end{figure}

The KLOE detector (fig.~\ref{fig:kloesec}) 
consists of a large cylindrical drift chamber surrounded by a lead-scintillating fiber
sampling calorimeter. A superconducting coil outside the calorimeter 
provides a 0.52\T\ field. The drift chamber~\cite{DCnim}, 
4\m\ in diameter and 3.3\m\ long, has 12\,582 all-stereo sense wires and 37\,746 
aluminium field wires. 
The chamber shell is 
made of carbon fiber-epoxy composite and the gas used is a 90\% helium, 10\% isobutane mixture. These 
features maximize transparency to photons and reduce \kl$\rightarrow$ \ks\ regeneration. 
The position resolutions are
$\sigma_{xy}\!\sim\!150\,\um$ and $\sigma_z\!\sim\!2\mm.$ The momentum resolution is 
$\sigma(p_{\perp})/p_{\perp}\!\leq\!0.4\%$. Vertices are reconstructed with a spatial 
resolution of $\sim\!3\mm$. 
The calorimeter~\cite{EmCnim} is divided into a barrel and two endcaps and covers 98\% of the 
solid angle. The energy resolution is $\sigma_E/E\!=\!5.7\%/\sqrt{E (\GeV)}$ and the timing 
resolution 
is $\sigma_t\!=\!54\ps/\sqrt{E (\GeV)}\oplus50\ps.$ The trigger~\cite{TRGnim} uses 
calorimeter and chamber 
information. For the work described here, the trigger relies entirely on 
calorimeter information. Two local energy deposits above threshold 
($50$\MeV\ on the barrel, $150$\MeV\ on the endcaps) are required.
The trigger time has a large spread with respect to the bunch crossing time. However, it is 
synchronized with the machine RF divided by 4, $T_{\rm sync}\!=\!10.8\ns$, with an accuracy 
of 50\ps. The time \tzero\ of the bunch crossing producing an event is determined after 
event reconstruction. 

\section*{The KLOE data set}
During 2002 data taking \DAF\  reached a peak luminosity of $\simeq 8
\times 10^{31} cm^{-2}s^{-1}$.
Since the start of data taking, we have collected 
an integrated luminosity of $\sim 500~$\Lpb ; considering only the
$\sim 400~$\Lpb\ of 
data passing all quality cuts, with particular reference to machine-background conditions,
and using a $\phi$ cross-section of $\sim 3 \mu$b, we can 
estimate that our data set contains
$\sim 6\times 10^{8}~~K^{+}K^{-}$ pairs and
$\sim 4\times 10^{8}~~K_{L}K_{S}$ pairs.  
Therefore, for \Keltre , taking selection efficiencies into account,
we estimate that we have about a million events for each of the
semileptonic channels, which translates into a  statistical error of $\sim$10$^{-3}$
on the branching ratio measurements. 
Moreover, these data give a statistical contribution to the absolute error on the
slope of \ff\ of $\sim$10$^{-4}$ for both neutral and charged kaons; 
we note here that $\lambda_{0}$ must be measured
with an absolute error
of $\sim$10$^{-3}$ to reach a 1\% relative precision 
on the theoretical determination of $f^{K \pi}_{+}(0)$~\cite{BandT}.

Two peculiar characteristics of KLOE, are the tagging technique and the good resolution on
kaon momentum.
The tagging allows us 
to select clean kaon beams of \kpm\ or of \ksl , and 
to measure absolute branching ratios.
Thus the strategy for the selection of $K_{l3}$ decays is to tag using
one kaon of the pair, and to look for the desired semileptonic decay of the other.  

\begin{figure}
  \hbox to\hsize{\hss
    \includegraphics[width=4.0cm]{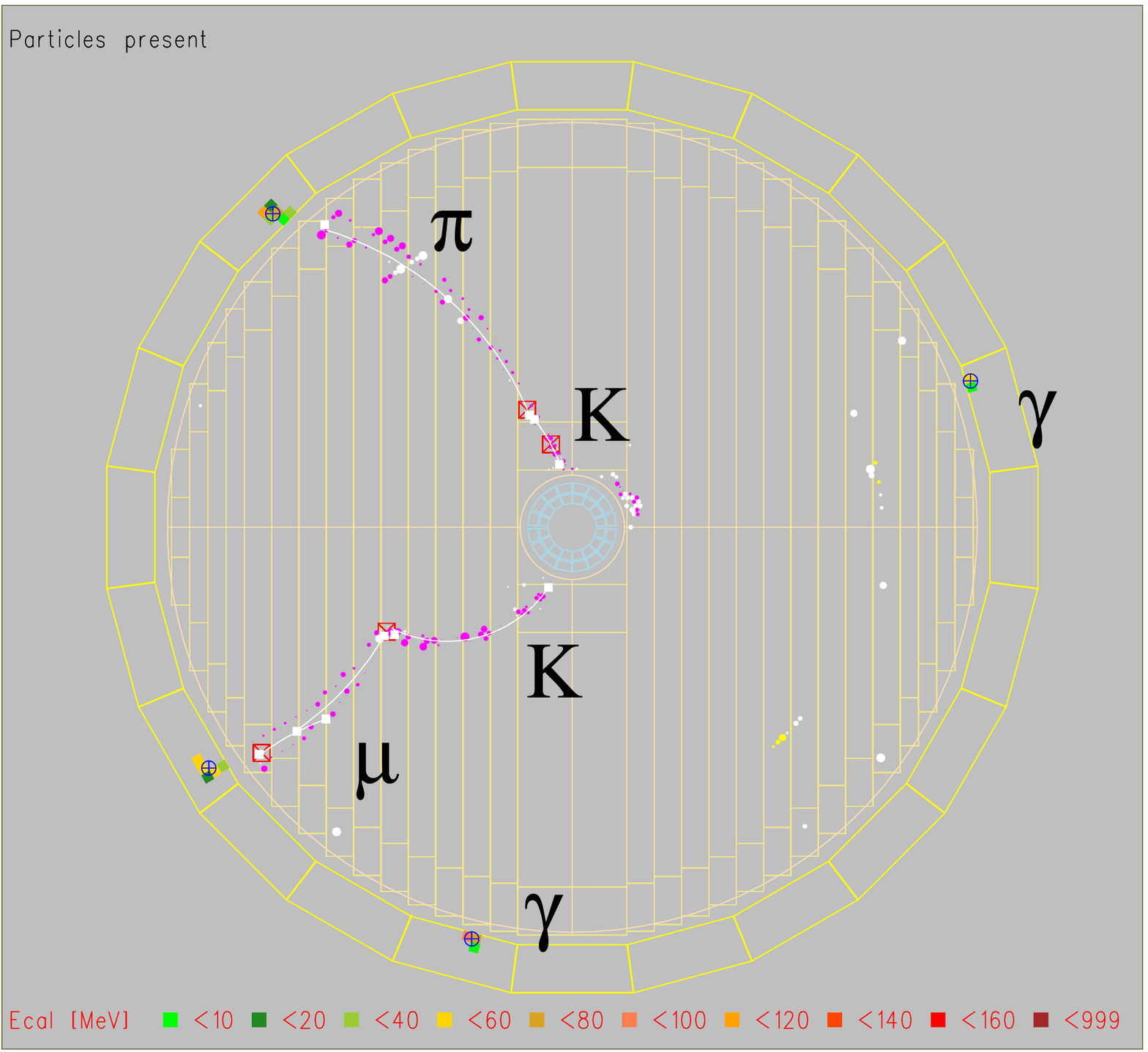}
    \includegraphics[width=3.4cm]{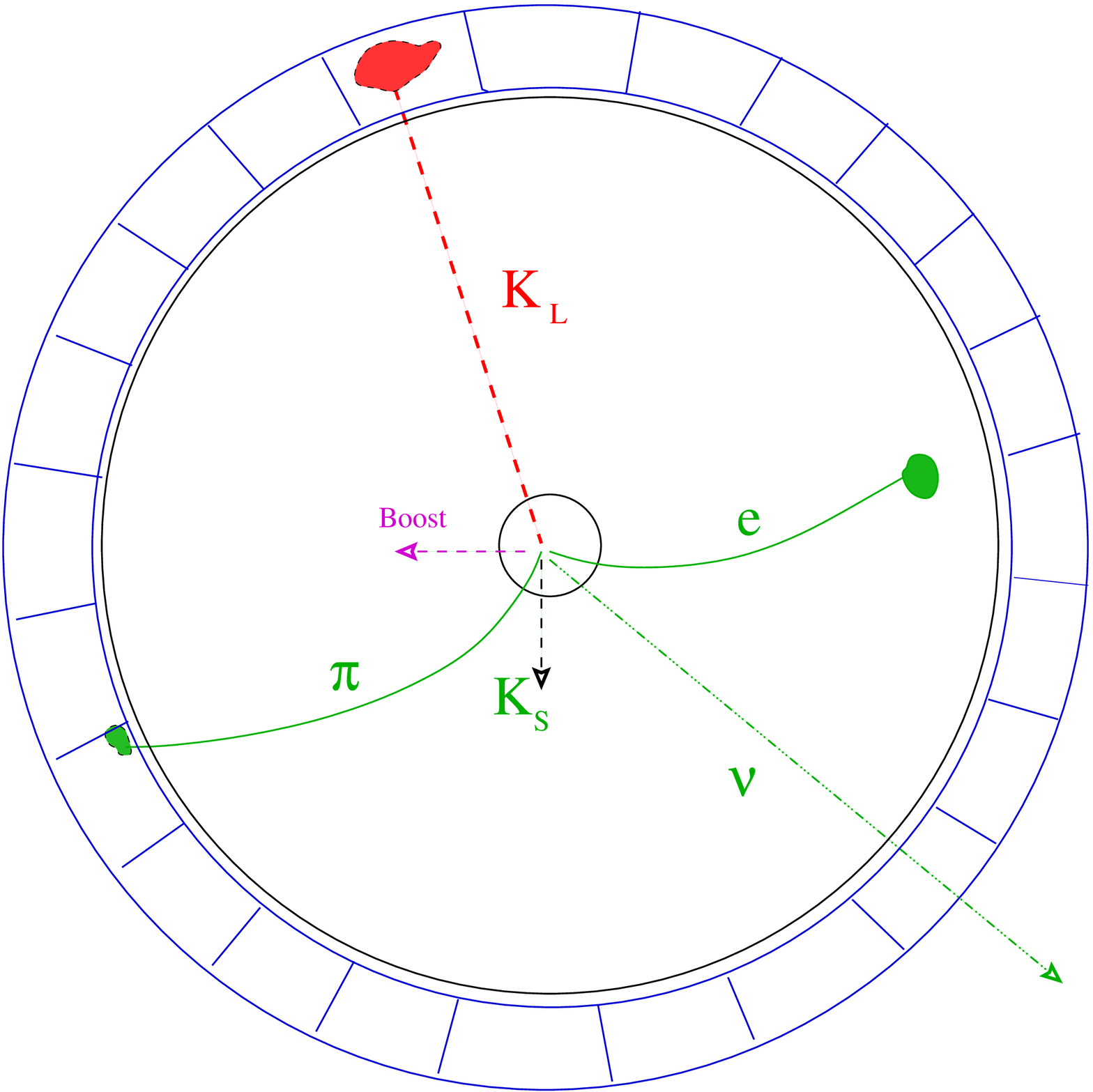}
    \hss}
  \caption{{\bf Left:} K$^{+}\to\pi\pi^{0}$ and K$^{-}\to\mu\nu$ event: the two decays used
    to tag K$^{\pm}$ events.  
    {\bf Right:} A \kl-{\it crash} and \DKSeIII\ event.}
  \label{fig:tagkpm}
\end{figure}

The very clean signature of the decays \kppo\ and \kmunu\ is
exploited to tag 
charged kaons~(fig.\ref{fig:tagkpm}, left); only drift 
chamber information is used to select these decays,
so that the tagging efficiencies 
for different decays of tagged particle
can be estimated directly from data, using the additional 
calorimetric information.

For neutral kaons, we use the \DKSpippim\ decays to tag the \kl , 
while the sample of \kl\ interacting in the calorimeter (\kl-{\it crash})
are used to tag the \ks~(fig.\ref{fig:tagkpm}, right).
In these cases as well the tagging efficiencies are estimated directly from data.
With the statistics of $\sim 400$~\Lpb\  we can
reach a level of ${\cal O}(0.1 \%)$ accuracy on tagging efficiencies,
both for charged and neutral kaons. 
 
As already mentioned,
the second peculiar characteristic of KLOE is the good resolution on the 
kaon momentum. For neutral kaons, the resolution on \kl\ momentum 
($p_L = p_\phi - p_S$) relies on the resolution
of the \ks\ momentum ($\simeq$1\MeV, measured with the drift chamber using \DKSpippim\ events),
and on the fact that $p_\phi$ is measured run by run
using Bhabha events and contributes a negligible error to the \kl\ resolution.
For charged kaons, the kaon momentum is measured directly by the drift chamber
with a resolution of $\simeq$1\MeV.

\section*{Selection of \Keltre\ samples}
\begin{figure}
  \hbox to\hsize{\hss
    \includegraphics[width=7.8cm]{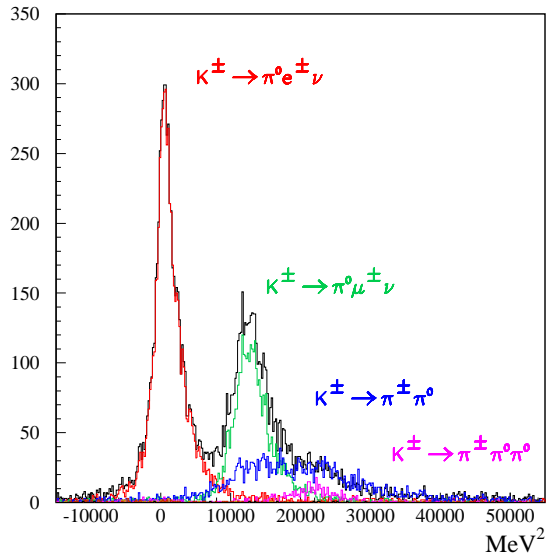}
    \hss}
  \caption{MC distribution of the $m^2$ of the charged daughter of the kaon,
    measured with a ToF technique: the \Kpmetre\ and \Kpmmutre\ peaks
    are clearly separated while a background from \kppo\ and \kppoo\ is
    still present.}
  \label{fig:ke3tof}
\end{figure}

For charged kaons, 
the sample of \Keltre\ events is selected by asking for a tag on one side, 
and for a decay vertex in the drift chamber and one $\pi^{0}$ in the electromagnetic
calorimeter on the other. 
The time of flight information can be used to separate
charged pions, muons, and electrons, exploiting the excellent timing resolution of
the detector~(fig.\ref{fig:ke3tof}). 
Most of the selection efficiencies can be evaluated directly from data 
using control samples: the momentum range of the lepton in $K^{\pm}_{l3}$ is
covered by \kppo\, \kmunu\ and \kppoo\ decays, while the energy range of 
\Keltre\ $\pi^{0}$ clusters is covered by $\pi^{0}$ clusters from
\kppo\ and \kppoo\ decays.

\begin{figure}
  \hbox to\hsize{\hss
    \includegraphics[width=7.8cm]{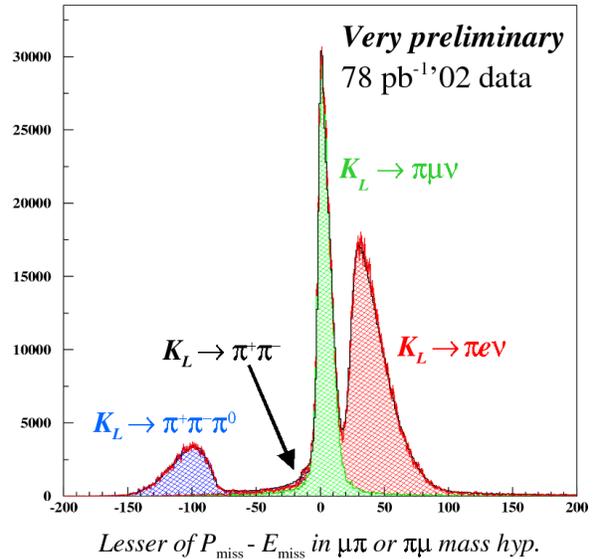}
    \hss}
  \caption{Distribution of the difference between
    missing momentum and missing energy (\MeV) for \kl\ decays to charged productd 
    in the kaon rest frame.}
  \label{fig:klcharged}
\end{figure}

A sample of semileptonic \kl\ decays 
has been selected from 78~\Lpb\ of '02 data
by looking for a \kl-{\it tag} on one side
and asking for a vertex in the drift chamber fiducial volume.
The separation of the various \kl\ decays is evident in the plot (fig.~\ref{fig:klcharged})
of P$_{miss}$--E$_{miss}$ in the $\mu\pi$ or $\pi\mu$ mass
(the mass assignment used is that which gives the smaller value of
P$_{miss}$--E$_{miss}$ in each event).
The difference of tagging efficiencies for the decay channel of the
tagged particle is less than 1\%, estimated by MC.
For this sample, fitting the data distribution in figure~\ref{fig:klcharged}
with the MC distribution, 
for each of the decay channels with independent normalization constants,
we get the branching ratios
\KOetre\ = $0.384 \pm 0.002$, \KOmutre\ = $0.271 \pm 0.002$, and 
$K^{0}\rightarrow\pi^+\pi^-\pi^0 = 0.132 \pm 0.002$, in good agreement
with PDG values~\cite{pdg02}.
These results are very preliminary: 
the errors are statistical only, and are dominated by the lack of MC statistics. 
The systematic errors are also at the 1-2\% level but have not yet been 
fully evaluated.
In the future, the time of flight information will be used to increase the
separation of pions from muons and electrons.

\section*{Kaon lifetime}
Looking at the PDG~\cite{pdg02} fit for \taupm , 
some discrepancies between ``in-flight'' and ``at-rest'' measurements can be found. 
Moreover, there are some differences between the various ``at-rest'' measurements 
obtained using different materials to stop kaons.
A new high-statistics measurement by KLOE could clarify 
this situation.

\begin{figure}
  \hbox to\hsize{\hss
    \includegraphics[width=7.8cm]{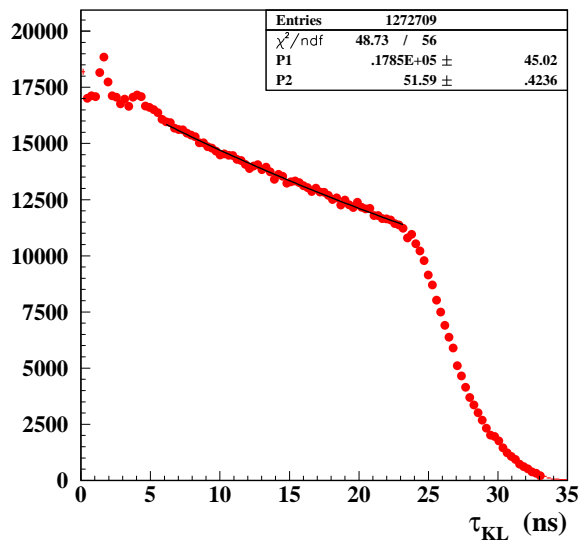}
    \hss}
  \caption{Proper time distribution of decays \kl $\rightarrow \pi^{0}\pi^{0}\pi^{0}$ events.}
  \label{fig:taul}
\end{figure}

For neutral kaons, KLOE has a preliminary measurement of the \kl\ lifetime 
as a by product of the measurement of the ratio 
$R = \Gamma(K_L \to \gamma\gamma) / \Gamma (K_L\to\pi^{0}\pi^{0}\pi^{0})$~\cite{gaia}.
The fit of the distribution of the \kl\ $\to\pi^{0}\pi^{0}\pi^{0}$ decay vertex position
(fig.~\ref{fig:taul}) gives $\tau = (51.5 \pm 0.5)$~ns 
(the error is statistical only, from the 36~\Lpb  used for this estimate),
in good agreement with the world average, $\tau = (51.7 \pm 0.4)$~ns~\cite{pdg02}).

\section*{Measuring $V_{us}$}
\begin{figure}
  \hbox to\hsize{\hss
    \includegraphics[width=\hsize]{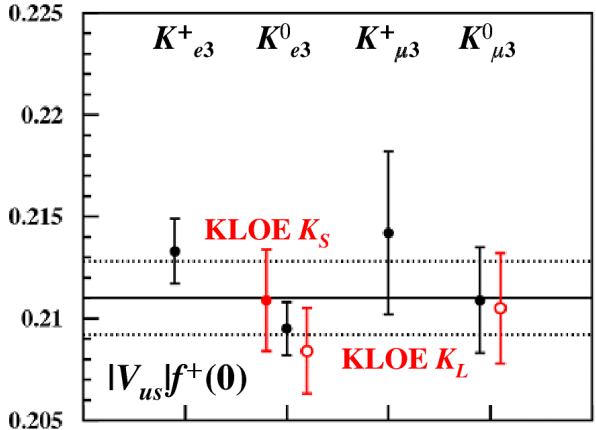}
    \hss}
  \caption{\Obs\ from the four \Keltre\ modes from PDG~\cite{pdg02} with the
    average over PDG electronic modes; in red the \Obs\ value for
    the KLOE branching ratio measurements.}
  \label{fig:vusnew}
\end{figure}

Using experimental inputs from the PDG~\cite{pdg02}, four evaluations of the 
physical observable \Obs\ can be made
(black points in fig.~\ref{fig:vusnew}). 
KLOE preliminary branching ratio measurements of \KOetre\ and \KOmutre\ 
confirm the previous value for \Obs\ 
(red open points in fig.~\ref{fig:vusnew}).
Also the KLOE preliminary branching ratio measurement of 
\DKSeIII\ = $(6.81 \pm 0.12 \pm 0.10)\times 10^{-4}$
(upgrade with 2001 data of the 2000 result~\cite{kssemil}) gives a \Obs\ value
in agreement with the KLOE \kl\ and PDG estimations 
(red point in fig.~\ref{fig:vusnew}). 
It must be noted that each KLOE point in fig.~\ref{fig:vusnew} has a statistical
precision comparable to the PDG one which represents the fit or average of several
experiments. Moreover, for most
of the branching ratio measurements in the PDG it is not clear
if these correspond to photon inclusive widths.

\section*{Conclusions}
The KLOE preliminary measurements of \kl\ charged decays and of \DKSeIII\ branching ratios are
currently at 2\% level, while the \taul\ is measured with a precision of 1\%.
Working is in progress for the other measurements involved in \Vus.
By measuring with the same detector the
absolute branching ratios and the form factors momentum dependence both for 
charged and neutral kaon semileptonic decays, KLOE can improve the situation 
of the CKM-matrix element \Vus .

% \section*{Acknowledgements}
% We thank the \DAF\ team for their efforts in maintaining low background running 
% conditions and their collaboration during all data-taking. 
% We want to thank our technical staff: 
% G.F.Fortugno for his dedicated work to ensure an efficient operation of the KLOE Computing Center; 
% M.Anelli for his continous support to the gas system and the safety of the detector; 
% A.Balla, M.Gatta, G.Corradi and G.Papalino for the maintenance of the electronics;
% M.Santoni, G.Paoluzzi and R.Rosellini for the general support to the detector; 
% C.Pinto (Bari), C.Pinto (Lecce), C.Piscitelli and A.Rossi for
% their help during shutdown periods.
% This work was supported in part by DOE grant DE-FG-02-97ER41027; 
% by EURODAPHNE, contract FMRX-CT98-0169; 
% by the German Federal Ministry of Education and Research (BMBF) contract 06-KA-957; 
% by Graduiertenkolleg `H.E. Phys. and Part. Astrophys.' of Deutsche Forschungsgemeinschaft,
% Contract No. GK 742; by INTAS, contracts 96-624, 99-37; 
% and by TARI, contract HPRI-CT-1999-00088. 

\end{document}